\documentclass{ws-mpla}

\renewcommand{\draftnote}{ }
\renewcommand{\trimmarks}{ }
\begin{document}

\markboth{D. Ebert, R. N. Faustov \& V. O. Galkin}
{Hyperfine Splitting and Leptonic Decay Rates} 

%
%

\title{\vspace{-1.5cm}
    \begin{flushright}
      \textmd{ HU-EP-03/15}
    \end{flushright}\vspace{1cm}
HYPERFINE SPLITTING AND LEPTONIC DECAY RATES\\ IN HEAVY QUARKONIA} 

\author{D. EBERT}

\address{Institut f\"ur Physik, Humboldt--Universit\"at zu Berlin,
  Newtonstr.15, D-12489 Berlin, Germany\\
  debert{@}physik.hu-berlin.de} 

\author{R. N. FAUSTOV$^*$ and V. O. GALKIN$^\dag$}

\address{Institut f\"ur Physik,
  Humboldt--Universit\"at zu Berlin, Newtonstr.15, D-12489 Berlin, 
  Germany\\ and\\
Russian Academy of Sciences, Scientific Council for
Cybernetics, Vavilov Str. 40, Moscow 117333, Russia\\
$^*$faustov@theory.sinp.msu.ru,  $^\dag$galkin@physik.hu-berlin.de}

\maketitle


\begin{abstract}
The hyperfine splitting in heavy quarkonia is considered in its
relation to the leptonic decay rates with the account of relativistic
and  radiative corrections. The calculated decay rates agree well with
the available experimental data, while the predicted $\eta_c(2S)$ mass
is significantly smaller than the value measured recently by the Belle
Collaboration. 

\end{abstract}

\ccode{PACS Nos.: 13.20.Gd,14.40.Gx, 12.39.Ki}

\vspace*{12pt}
\noindent
Recent observation of the charmonium $2^1\!S_0$ state $\eta_c(2S)$ by
the Belle Collaboration yielded the following values for its mass:
\begin{equation}
  \label{eq:bv}
  M(\eta_c(2S))=\left\{
      \begin{array}{c}
3.654(14)\ \textrm{GeV  (see Ref. 1)} \cr
3.622(12)\ \textrm{GeV  (see Ref. 2)}
      \end{array}\right. .
\end{equation}
These values are significantly larger than most predictions of the
constituent quark models\cite{efg,eq} and the previous (unconfirmed)
experimental value\cite{cbc} $M[\eta_c(2S)]=3.594(5)$~GeV. The resulting $2S$
hyperfine splitting (HFS) would be about 2--4 times smaller than the
$1S$ HFS in charmonium which is quite unexpected\cite{eq,elq,bb}. The HFS
is closely connected with leptonic decay rates and vector decay constants of
heavy quarkonia. This connection for charmonium was discussed at
length in the recent paper\cite{bb}. Here we extend this discussion in
order to include the relativistic corrections for the vector
constants\cite{cea}. We considered these corrections for heavy-light
($B$ and $D$) mesons in our recent paper\cite{fconst}.

The vector decay constant $f_V$ is defined as follows
\begin{equation}
  \label{eq:fcv}
  \left<0|\bar Q\gamma^\mu Q|V({\bf K},\varepsilon)\right>=f_V
  M\varepsilon^\mu,
\end{equation}
where ${\bf K}$ is the quarkonium momentum, $\varepsilon$ and $M$ are
the polarization vector and mass of the quarkonium. The relativistic
expression for $f_V$ can be obtained from Eq.~(3) of
Ref.~\refcite{fconst} by putting $m_q=m_Q=m$   
\begin{equation}
  \label{eq:fv}
 f_V=\sqrt{\frac{12}{M}}\int\frac{d^3p}{(2\pi)^3}
 \left\{1-\frac{\epsilon(p)-m}{3\epsilon(p)}\right\}\Phi_V(p), 
\end{equation}
where $\epsilon(p)=\sqrt{{\bf p}^2+m^2}$ and $\Phi_V(p)$ is the vector
quarkonium wave function in the momentum space.

In the nonrelativistic limit $p^2/m^2\to 0$ this expression reduces to the
well-known formula
\begin{equation}
  \label{eq:fnr}
  f_V^{\rm NR}=\sqrt{\frac{12}{M}}\left|\Psi_V(0)\right|,
\end{equation}
where $\Psi_V(0)$ is the wave function at the origin $r=0$.

The leptonic decay rate for zero lepton mass is given by the following
relation
\begin{equation}
  \label{eq:gamma}
  \Gamma_0(V\to e^+e^-)=\frac{4\pi\alpha^2e_Q^2}{3M}f_V^2,
\end{equation}
where $\alpha$ is the QED fine structure constant and $e_Q$ is the
quark charge in units of the elementary electric charge.  Using
(\ref{eq:fnr})  we obtain in the nonrelativistic limit the widely-used
relation
\begin{equation}
  \label{eq:gnr}
  \Gamma^{\rm{NR}}=\frac{16\pi\alpha^2e_Q^2}{M^2}
  \left|\Psi_V(0)\right|^2. 
\end{equation}

The one-loop QCD corrections modify Eq.~(\ref{eq:gamma}) in the
following way\cite{bgkk}
\begin{equation}
  \label{eq:grc}
  \Gamma=\Gamma_0\left(1-\frac{16}{3\pi}\alpha_s\right),
\end{equation}
where we take the QCD coupling constant $\alpha_s$ according to
PDG\cite{pdg} equal to
\[
\alpha_s(m_c)=0.26,
\qquad  \alpha_s(m_b)=0.18.
\]

\begin{table}[h]
\tbl{Heavy quarkonium masses (in GeV) and HFS $\Delta M$ (in
  MeV) of $1S$ and $2S$ states.} 
{\begin{tabular}{@{}ccccccc@{}} \toprule
States& $1^1\!S_0$&$1^3\!S_1$ &$\Delta M^{\rm{HFS}}_{1S}$&
$2^1\!S_0$&$2^3\!S_1$ &$\Delta M^{\rm{HFS}}_{2S}$ \\\colrule
$c\bar c$& $\eta_c(1S)$& $J/\psi(1S)$& &$\eta_c(2S)$ &$\psi(2S)$& \\
theory\cite{efg}&2.979& 3.096& 117& 3.588& 3.686& 98\\
experiment\cite{pdg}&2.9797(15)&3.09687(4)&117(1)  & & 3.68596(9)&\\
experiment\cite{belle1}&2.979(2)&    & 117(1)      &3.654(14)&
&32(14)\\ 
experiment\cite{belle2}&2.979(2)&    & 117(1)      &3.622(12)&
&64(12)\\
experiment\cite{cbc}& &   & & 3.594(5)&  & 92(5)\\
\colrule
$b\bar b$&$\eta_b(1S)$& $\Upsilon(1S)$& &$\eta_b(2S)$ &$\Upsilon(2S)$&
\\
theory\cite{efg}& 9.400& 9.460 & 60& 9.993 & 10.023& 30\\
experiment\cite{pdg}& & 9.46030(26)& & &10.02326(31)
\\ \botrule
\end{tabular}}
\end{table} 

The heavy quarkonium mass spectra including all relativistic $v^2/c^2$
and one-loop radiative corrections were calculated in
Ref.~\refcite{efg}. The masses $M$ and HFS $\Delta
M^{\rm{HFS}}_{nS}\equiv M(n^3\!S_1)-M(n^1\!S_0)$ of $nS$  states
($n=1,2$) are presented in Table~1.

The ratio of the $2S$ to $1S$ HFS in charmonium is equal to
\begin{equation}
  \label{eq:r}
  R\equiv\frac{\Delta M^{\rm HFS}_{2S}}{\Delta M^{\rm HFS}_{1S}}=
  \left\{\begin{array}{ll}  
0.84(9) & \textrm{Ref. 3}\\
0.67(10)&\textrm{Ref. 4}\\
0.59(8)\eta_{\rm HF}, \  \eta_{\rm HF}\sim 1 \ & \textrm{Ref. 7}\\
0.6(2)  &\textrm{Ref. 10}
\end{array}\right.
\end{equation}
The upper high value of $R$ is obtained in the framework of the
relativistic quark model\cite{efg}. The bound state equation used
there takes into account the one-loop radiative corrections to
the static potential enlarging the effective Coulomb constant
$\alpha_V$ and relativistic effects both increasing the value of $R$.

Inserting into Eq.~(\ref{eq:fv}) the same wave functions as used in
Ref.~\refcite{efg} we can calculate on the basis of
Eqs.~(\ref{eq:gamma}) and (\ref{eq:grc}) the leptonic decay rates of
vector quarkonia. The results of these calculations in comparison with
experimental data are shown in Table~2.

\begin{table}[h]
\tbl{Vector decay constants and leptonic decay rates of vector quarkonia.}
{\begin{tabular}{@{}ccccc@{}} \toprule
Decay&$f_V$&$\Gamma^{\rm NR}$&$\Gamma$&$\Gamma^{\rm exp}$\\
modes& MeV& keV&  keV& keV\\ \colrule
$J/\psi(1S)\to e^+e^-$ &551 & 6.7 & 5.4 &5.26(37)\\
$\psi(2S)\to e^+e^-$ &401  & 3.2 & 2.4 &2.19(15)\\
$\Upsilon(1S)\to e^+e^-$ &839 & 1.4 & 1.3 & 1.32(5)\\
$\Upsilon(2S)\to e^+e^-$ &562 & 0.6 & 0.5 & 0.52(3)
\\ \botrule
\end{tabular}}
\end{table} 

As seen from Table~2, relativistic corrections considerably reduce the
calculated decay rates and bring them in good agreement with
experimental values. Thus we observe the overall selfconsistency
between predictions for HFS and leptonic decay rates of heavy
quarkonia. It means that the new Belle data\cite{belle1} for the
$\eta_c(2S)$ mass cannot be easily explained in the framework of
relativistic constituent quark models and, if confirmed, require some
novel ideas and approaches (scale choice\cite{bb} for $\alpha_s$, state
mixing\cite{mr}, etc.).

\section*{Acknowledgments}
The authors express their gratitude  E. Eichten, A. Martin,
M. M\"uller-Preussker and V.~Savrin   
for support and discussions. Two of us (R.N.F. and V.O.G.) 
were supported in part by the {\it Deutsche
Forschungsgemeinschaft} under contract Eb 139/2-2.

\end{document}